\documentclass[conference]{IEEEtran}
\usepackage{graphicx}
\usepackage{amsmath}
\usepackage{tabularx}
\usepackage{rotating}
\usepackage{amssymb}
\usepackage{subfig}
\usepackage{amsthm}
\usepackage{soul, color}

\usepackage[linesnumbered,ruled,vlined]{algorithm2e}
\usepackage[linesnumbered]{algorithm2e}

\newtheorem{problem}{Problem}

\def\IEEEbibitemsep{0pt plus .5pt}
\begin{document}

\title{Optimal Multicasting in Hybrid RF/FSO DTNs}
\author{\IEEEauthorblockN{Michael Atakora and Harsha Chenji}%
\IEEEauthorblockA{School of Electrical Engineering and Computer Science, Ohio University, USA\\ ma136412, chenji@ohio.edu\vspace*{-1em}}}%
\maketitle

\begin{abstract}
The multi-copy routing paradigm in Delay Tolerant Networks (DTNs) implies that increasing contact bandwidth leads to a decrease in data delivery delay and an improvement in throughput. With Hybrid Radio Frequency/Free Space Optical (RF/FSO) PHY layers, the high data rate FSO links can be used to increase the contact bandwidth. However, due to the highly directional nature of FSO links, broadcasting is difficult. A na\"{i}ve broadcast strategy where the beam divergence angle is increased to include many nodes in the broadcast set results in low data rate, and does not always result in the minimum achievable delay.

In this work we develop an optimal multicast algorithm for hybrid RF/FSO networks. We show that the problem is an abstraction of the minimum weight set cover problem which is known to be NP-hard. A computationally cheap greedy local optimum heuristic is proposed. A comprehensive evaluation using delay, throughput and computation time as metrics is performed using various solutions. These extensive evaluations show that our solution outperforms both na\"{i}ve broadcast and multiple unicast, taking 95\% less time as compared to the exact algorithm, while providing comparable performance.
\end{abstract}

\IEEEpeerreviewmaketitle
\pagestyle{plain}
\thispagestyle{plain}

\section{Introduction}
Radio Frequency (RF) spectrum congestion, as well as an increase in mobile data traffic has led to a search for alternate PHY layers. Over the last few decades, systems which use free space optics (FSO) operating in the 352-384 and 187-197 THz ranges~\cite{majumLaser} have been developed. Recently, the NASA LLCD project demonstrated a 622 Mbps optical link from Earth to Moon~\cite{borosonLLCD}. The spate of recent advances in visible light communication has reinforced this trend. However, FSO is highly directional, and suffers from drawbacks such as requiring near perfect alignment of both transmitter and receiver, as well as susceptibility to absorption in the atmosphere. 

Delay tolerant networks (DTNs)~\cite{Fall03} cover large geographical areas in which node density is sparse but node mobility is high (e.g., space networks). Node inter-contact opportunities are limited due to low node density and low (compared to the size of the deployment area) transmission range. Therefore, for successful delivery of packets, DTN protocols rely on multi-copy routing~\cite{Liu2011} and store-carry-forward approaches. By replicating a packet onto multiple nodes, the packet delivery probability is increased. Examples of such protocols are Epidemic~\cite{epidermic} and Spray and Wait~\cite{sprayWait}. The number of nodes onto which packets are replicated during a contact opportunity is proportional to the contact bandwidth, and it is well known~\cite{throwBox} that the contact bandwidth influences the performance of a DTN. Therefore, any DTN routing protocol will benefit from increased contact bandwidth. 

The use of high bandwidth FSO (or hybrid RF/FSO) links in DTNs has been proposed before~\cite{nichols07}, and similarly, directional RF~\cite{nichols08}. While FSO has high data rates as compared to RF, it also highly directional, which makes broadcasting or multicasting inefficient. Consequently, the contact bandwidth reduces and the DTN suffers a performance hit. In this paper, we investigate how a PHY layer ``knob'' that is unique to FSO can be used to increase the contact bandwidth. Visible light communication PHY radios such as LEDs~\cite{disneyVLC} have divergence angles (defined in Section~\ref{sysMod})  in the tens of degrees range, while those used by NASA are measured in microradians~\cite{borosonLLCD}. As we will show, simply increasing the divergence angle to include many nodes in the broadcast set will result in a reduction of data rate. At the same time, performing a ``multiple unicast'' is not optimal since the time taken to re-align the transmitter and receiver after each transmission is non-zero. Therefore, there is an inherent \textit{tradeoff} between increasing the divergence angle to cover multiple nodes, versus unicasting to multiple nodes one after the other.

In this paper, we investigate how optimal multicasting can be performed in DTNs with nodes equipped with both RF and FSO radios. The RF radio, which is assumed to be orders of magnitude slower than the FSO radio, is used as a control channel for the FSO link. After accounting for uncertainty in node location, the source node in question calculates an \textit{optimal multicast set} (which defines several consecutive data transmissions) such that the data delivery delay is minimized. We show that the problem is equivalent to a weighted set cover problem. A known solution based on integer linear programming is proposed, as well as a greedy heuristic. Finally, we evaluate the performance of our solution. The rest of this paper is laid out as follows: we motivate the need for our research and review related work in Section \textrm{II}. In Section \textrm{III}, we present the optimal multicast algorithm. A performance evaluation of the optimal solution, a heuristic and other schemes is presented in Section \textrm{IV}, after which we provide a conclusion.

\section{Motivation \& Related Work}
In this section we first motivate our work by emphasizing the importance of increasing the number of contact opportunities in DTNs, and by demonstrating the viability of FSO communications. Next, we place our work in context by discussing recent research related to optimal multicasting - especially in networks with directional RF radios.

\textit{Motivation:-} It has been shown in~\cite{throwBox} that increasing the number of contact opportunities results in lower delay and higher throughput. Mobility further reduces contact duration, therefore using a PHY with high effective data rates (e.g FSO) is desirable. However, broadcasting data with FSO might not always be optimal, since a large divergence angle leads to low transmission rates. To the best of our knowledge, the optimal multicast problem in wireless optical ad hoc networks has not been addressed. Therefore, in this paper, we seek to address this problem. We develop an optimal strategy based on the minimum weighted set cover problem. In addition, we develop a faster heuristic, after which we compare the performance of the various schemes. 

\begin{figure}[t]
	\centering
	\subfloat[FSO Link]{\includegraphics[height=7em]{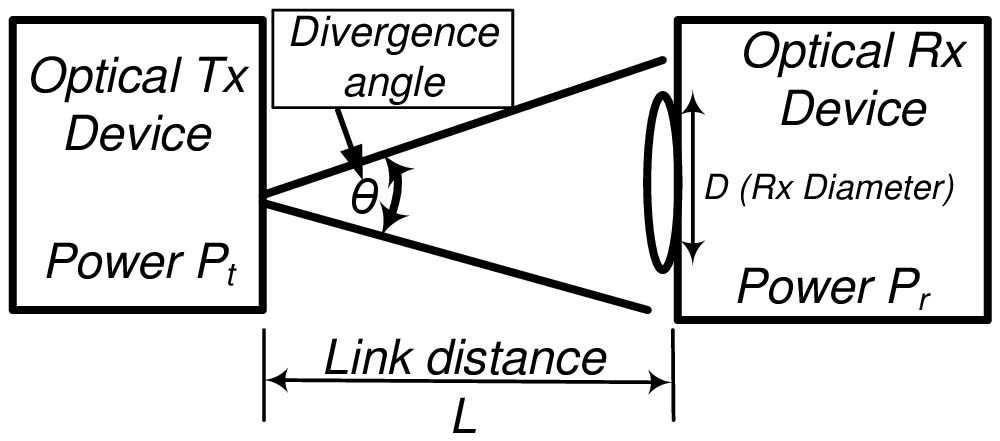}\label{fig:FSOlink}}\hspace*{1em}
	\subfloat[Hybrid RF/FSO Net.]{\includegraphics[height=8em]{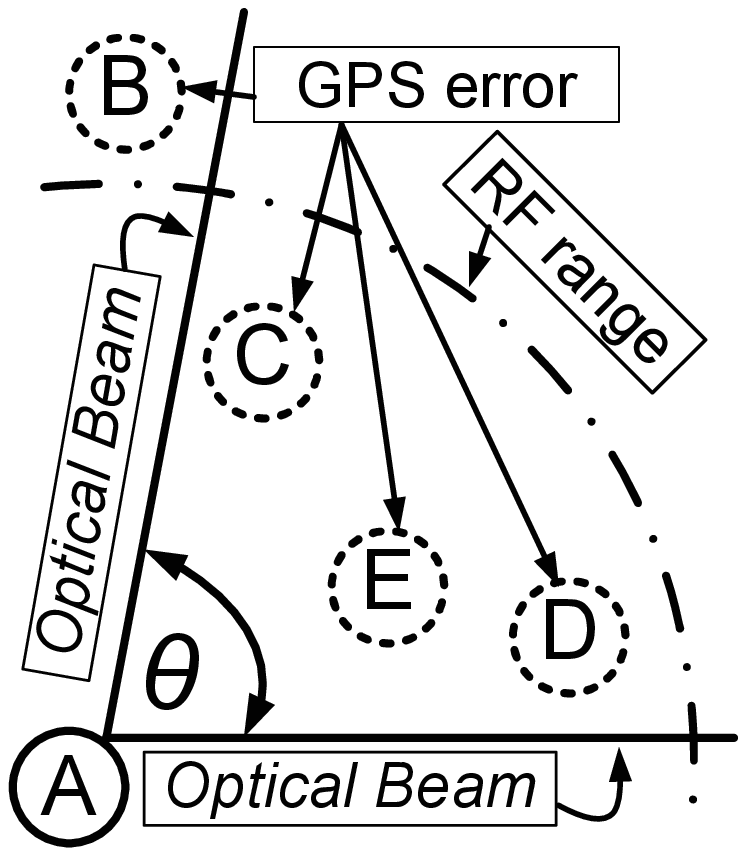}\label{fig:networkDiag}}
	\caption{(a) A hybrid RF/FSO network where node A communicates with nodes C, D \& E with a FSO beam divergence angle of $\theta$. Node A is unaware of node B since it is outside of node A's RF radio range. (b) Illustrating the various FSO link parameters for a single point-to-point link: $P_t, P_r, L, \theta$ and $D$.}
	\label{fig:sysmodel}
\end{figure}

\textit{Related Work:-} Multicast and broadcast algorithms for ad hoc networks with directional RF antennas have been investigated by a number of researchers. In~\cite{directAntWie02, directAntWie01, directAntWie02Infocom}, the authors develop and evaluate broadcast and multicast heuristics for a network consisting of power constrained devices using tree construction algorithms. A multi-hop relaying scheme was employed in~\cite{lifetimeHou07, lifetimeHou05} to develop an online heuristic for multicast routing  with the objective of reducing energy consumption and prolonging network lifetime. Particle swarm optimization is used in~\cite{psoDelay} to develop a multicast algorithm with delay constraints. These efforts however do not address wireless optical broadcast. In the area of ad hoc networks in which nodes have FSO capability, a lot of research has been conducted. In~\cite{Yuksel2007, protoSevinc10, multiElBilgi08, bilgiMultTrans2010} the authors introduce and implement a FSO node design in which spherical surfaces are tessellated with several transceivers to achieve near omnidirectional node coverage. They achieve this by means of an auto-alignment circuit that detects a loss of line of sight by electronically tracking optical beams. The research efforts in~\cite{Yuksel2007, protoSevinc10, multiElBilgi08, bilgiMultTrans2010} do not address ad hoc networks in which nodes have both RF and FSO transmitters. In addition, they do not adapt beam divergence. In the area of FSO DTNs, the authors of~\cite{nichols07, nichols07perf, nichols08} develop algorithms for networks with fragile links. The objective of such algorithms is the minimization of a transient information level metric defined to be a function of both the amount of information in the network and the projected physical distance to the destination, where constraints such as QoS, varied network traffic, transmission and storage limits are incorporated into the algorithm.	

The work we present in this paper is different from previous research efforts in the area of multicast in directional RF ad hoc networks, omnidirectional FSO ad hoc networks, and hybrid RF/DTNs. To the best of our knowledge, this paper is the first to address the optimal multicast problem in hybrid RF/FSO DTNs. 

\section{Optimal Multicast Algorithm}
In this section we present our optimal multicast algorithm. First, the system model is defined. Then, the problem formulation is posed. We then show that the problem is equivalent to the minimum weighted set cover, which is NP-hard. Two solutions are proposed: a slower integer programming solution which achieves the best known solution, and a fast greedy-based heuristic. 

\subsection{Background \& System Model}\label{sysMod}
In free space optical communication, photons are generated at the source, and are collected by a receiver at the destination. Modulation is accomplished by either modulating multiple bits onto each photon~\cite{bitsPerPhoton}, or through photon counting. In the latter, the receiver records a binary \texttt{1} only when a certain number of photons are counted in a time period. The light beam that is generated either diverges due to physical imperfections in the source, or can be made to diverge using a lens; this angle of divergence is defined as the \textit{beam divergence angle}.

An FSO link is depicted in Figure~\ref{fig:FSOlink}. The beam divergence angle is denoted as $\theta$ (in radians), and $L$ is the Euclidean distance (in meters) between the sender and receiver. Given these parameters, the received power $P_{r}$ is expressed~\cite{majumLaser} as
\begin{equation}
P_r = {P_t} {\left(\frac{D}{{{\theta}}{L}} \right)^2}{L_{tp}}{L_{rp}}{\eta_t}{\eta_r}{10^\frac{{-\alpha}L}{10^4}}\label{linkPower}
\end{equation}
where $P_{t}$ is the transmitted power in watts, $D$ is the receiver diameter in meters, $L_{tp}$ and $L_{rp}$ are the pointing losses resulting from imperfect alignment of the transmitter and receiver respectively, $\eta_t$ and $\eta_r$ are the transmitter and receiver optical efficiencies respectively, and $\alpha$ is the atmospheric attenuation factor in dB/km. The photodetector sensitivity $N_{b}$ (in photons per bit) is the number of photons required to register a binary \texttt{1} at a specified bit error rate. With a light source of frequency $f$, the effective data rate $R_b$ at a divergence angle of $\theta$ is:

\begin{equation}
R_b(\theta) = \frac{P_{r}}{hf{N_{b}}} =\frac {{P_t}{D^2}{L_{tp}}{L_{rp}}{\eta_t}{\eta_r}{10^\frac{{-\alpha}L}{10^4}}}{hf{N_{b}}{{\theta}^2}{{L}^2}}\label{eqn:FSODataRate}
\end{equation}
where $h$ is Planck's constant. 

Based on the above theory, a system model is developed, and is shown in Figure~\ref{fig:sysmodel}. We assume that nodes (``A,B,C,D,E" in Figure~\ref{fig:networkDiag}) are equipped with an omnidirectional RF radio as well as a directional FSO radio. Nodes are able to obtain their position, and they broadcast it periodically using the RF radio. In addition to the dissemination of control information, RF links also serve as backup to FSO data links, should the latter be inactive. We account for possible positioning errors (dotted line around nodes in Figure~\ref{fig:networkDiag}), motivated by the fact that GPS systems currently have a 3 $m$ position accuracy 95\% of the time. Therefore, the nodes have to set the beam divergence angle $\theta$ such that any error in localization does not lead to link misalignment. For a particular node, $\theta$ can be easily calculated by finding tangents to the circle around it, whose radius represents the location accuracy. Therefore, we assume that nodes outside the RF radio range (``B'' in Figure~\ref{fig:networkDiag}) are not neighbors w.r.t FSO since their location cannot be obtained. 

\begin{figure}[t]
	\centering
	\subfloat[Broadcast]{\includegraphics[height=5em]{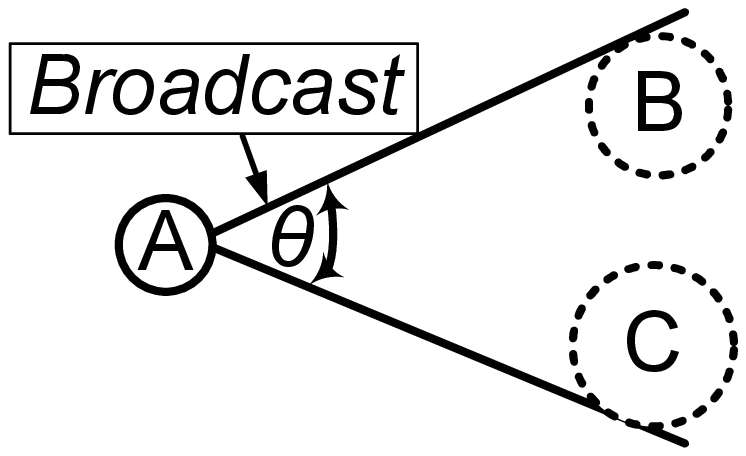}\label{fig:brod}}\hspace*{3em}
	\subfloat[Multiple unicast]{\includegraphics[height=5em]{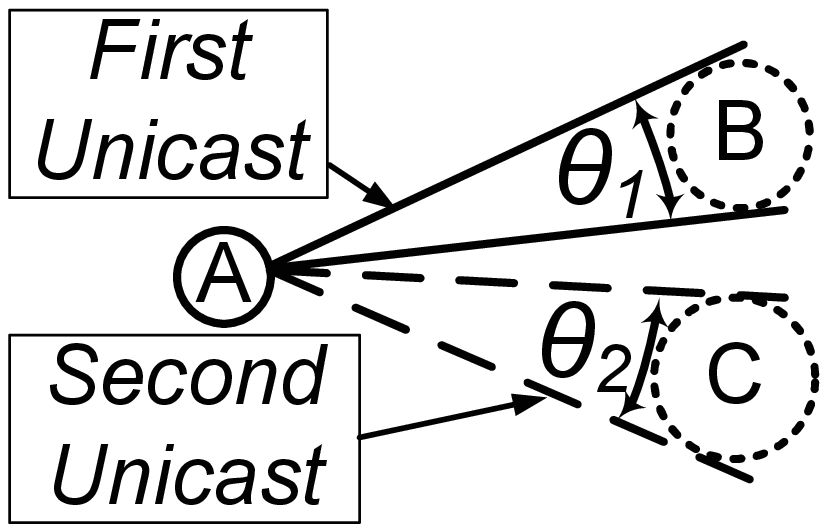}\label{fig:mulUni}}
	\caption{Illustrating the difference between (a) broadcast with a FSO beam divergence angle of $\theta$, and (b) multiple (two) unicasts with divergence angles of $\theta_1$ and $\theta_2$. Node A is able to deliver data to nodes B \& C in both cases, but since $\theta > \theta_1 + \theta_2$, aggregate $R_b$ in both cases might not be equal.}
	\label{fig:brodUnicast}
\end{figure}

\subsection{Broadcast versus Multiple Unicast}\label{brodMU} 
The broadcast primitive can be implemented in FSO by changing a node's divergence angle (Figure~\ref{fig:brodUnicast}). For example, node A intends to broadcast data to nodes B \& C. Node A can deliver data to B \& C simultaneously by increasing its $\theta$ such that B \& C lie within the beam (Figure~\ref{fig:brod}). However, the data rate $R_b$ of optical wireless links varies inversely with the square of the beam divergence angle (Equation~\ref{eqn:FSODataRate}). Therefore, in general, arbitrarily increasing $\theta$ to include all nodes will result in high transmission delay due to low data rates. One might think that to reduce the total broadcast delay, multiple unicast (Figure~\ref{fig:mulUni}) might be a viable alternative. Node A first transmits to B using $\theta = \theta_1$, and subsequently to C with $\theta = \theta_2$. Clearly, $\theta > \theta_1 + \theta_2$, and therefore, $R_b(\theta)$ is in general not equal to $R_b(\theta_1) + R_b(\theta_2)$. Moreover, with multiple unicast, the sender has to always realign its laser transmitter after each transmission. Even though multiple unicast may result in low per-node transmission delays, achieving perfect alignment is challenging and introduces non-zero alignment delay. We define \textit{alignment delay} $d_{al}$ as the time it takes a node to perfectly reorient it's laser transmitter in the direction of another node. 

With the aforementioned challenges, a broadcast strategy which minimizes the data delivery delay is necessary. Such a strategy would group nodes in a manner that maximizes both the data rate for each group, as well as minimizes the number of groups (to avoid $d_{al}$). In formulating such a strategy, we first define a universe $\mathcal{U}$ of nodes that are to receive broadcast data. A \textit{set} $S_i$ is a group of nodes in the network whereby exactly one transmission is required to multicast to each of it's elements. In other words, within a set, the sender does not realign its optical transmitter. It is easily seen that the union of all sets $S_i$ should be equal to the universe $\mathcal{U}$.

\begin{figure}[t]
	\centering
	\subfloat[Set with all nodes]{\includegraphics[height=7em]{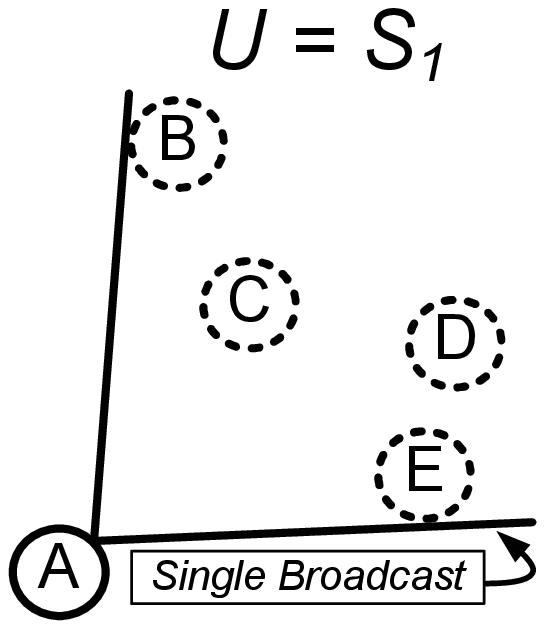}\label{fig:def3}}\hspace*{1.5em}
	\subfloat[A possible combination of sets ]{\includegraphics[height=7em]{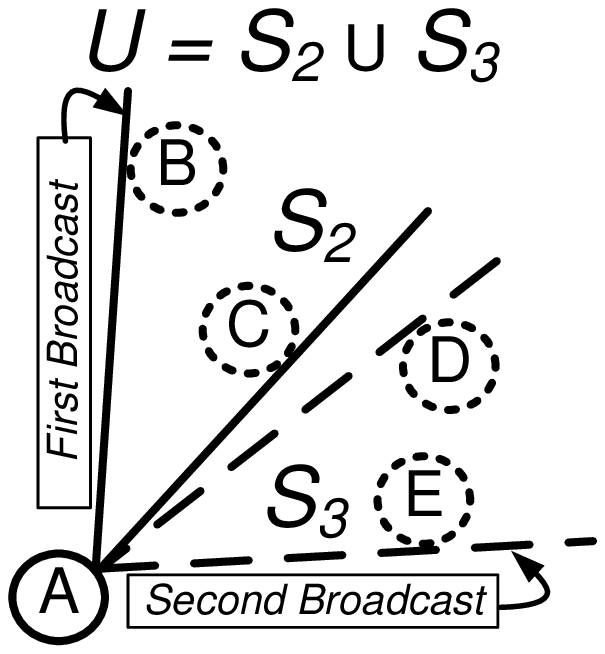}\label{fig:def4}}\hspace*{1.5em}
	\subfloat[Another combination of subsets]{\includegraphics[height=7em]{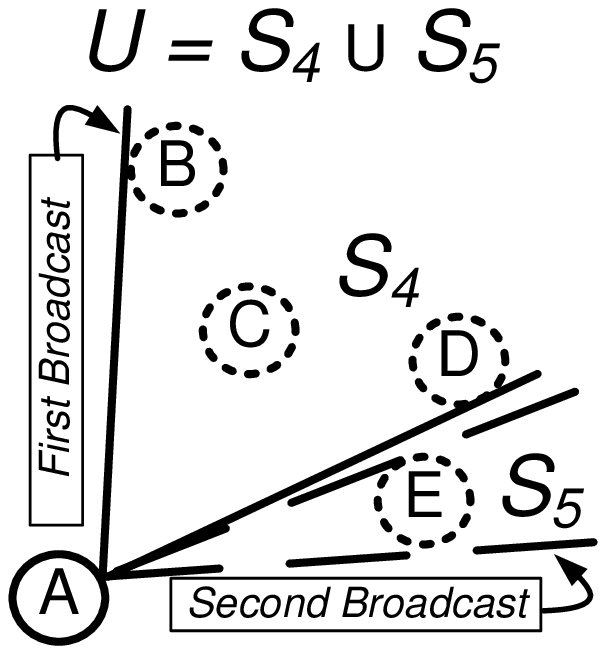}\label{fig:def5}}
	\caption{Illustrating several possible set combinations in FSO multicast (a) a single set (b) two sets with size of 2 each (c) two sets with size 3 and 1. The union of all sets in a diagram is always equal to the universe of nodes.}
	\label{fig:subset}
\end{figure}

We use Figure~\ref{fig:subset} to reinforce the notion of a universe and set. Here, node A broadcasts data to $\mathcal{U} = \{B, C, D, E\}$. There are different ways to build $\mathcal{U}$, by considering different combinations of sets $S_i$. In Figure~\ref{fig:def3}, a single set $S_1=\{B,C,D,E\}$ is sufficient. In Figure~\ref{fig:def4}, we use two sets $S_2 = \{B,C\}$ and $S_3=\{D,E\}$, each containing two elements. In Figure~\ref{fig:def5}, there are again two sets, one of size 3 ($S_4=\{B,C,D\}$) and the other of size 1 ($S_5=\{E\}$). We see from the three examples that while the union of sets is always equal to $\mathcal{U}$, the divergence angles (and hence the data rates) are different. An interesting property of this notion of sets in FSO is also highlighted. Supposing that we sort the nodes in order of decreasing azimuth $\phi$ from the origin (where node A is located). For three nodes $B,C,E$, if $\phi_E < \phi_C < \phi_B$, then any set containing nodes B and E should necessarily contain node C. This is the case in Figure~\ref{fig:def3}. A similar example is highlighted in Figure~\ref{fig:def5}, where a set containing nodes B and D also contains node C. This structure leads us to develop our problem formulation.

\subsection{Problem Formulation}\label{brodProb}
In enumerating all possible sets for a given $\mathcal{U}$ from Figure~\ref{fig:subset}, we see that there exactly 4 sets of cardinality 1 (not shown in Figure~\ref{fig:subset}):$\{B\},\{C\},\{D\}$ and $\{E\}$. Similarly, there are also exactly 3 sets of cardinality 2:$\{B,C\}, \{C,D\}, \{D,E\}$, exactly 2 sets of cardinality 3, and exactly 1 set of cardinality 4. Therefore generally, to broadcast to $N$ nodes, there are exactly $N$ sets of cardinality 1, exactly $N-1$ sets of cardinality 2,$\dots$, and exactly 1 set of cardinality $N$, for a total of $K = N(N+1)/2$.

The FSO optimal multicast problem can be stated as follows: given a universe $\mathcal{U} = \{n_1, n_2, \dots, n_N\}$ of $N$ nodes, a collection $\mathcal{S} = \{S_1, S_2, \dots, S_K\}$ of $K = N(N+1)/2$ sets can be constructed. The cost of broadcasting data to a set $S_i$ is the data delivery delay $d_i$ which depends on $R_b$ for that set, which in turn depends on the required $\theta$. The objective is to find $\mathcal{S'} \in \mathcal{S}$ with minimum total delay such that all $N$ nodes are covered. The delivery delay $d_i$ for a set $S_i$ is computed using the size of the broadcast data $P$, the minimum divergence angle $\theta_{i}$ required for all member nodes to be in the transmitter's footprint, and alignment delay $d_{al}$. Using Equation~\ref{eqn:FSODataRate}, $d_i$ is calculated as
\begin{equation}
d_{i} = \max\limits_{j}\left\{{\frac{P}{R_b(\theta_i)}} + {d_{al}}\right\} \text{ where } 1\leq j\leq |S_{i}|\label{eqn:subsetDelay}
\end{equation}
where $R_b(\theta_i)$ is calculated for each node $j \in S_i$ using different values of distance $L_j$. We formulate the optimal FSO multicast problem as a 0/1 integer problem. Each set $S_{i}$ is assigned a binary decision variable: $x_{i}$ is 1 if $S_{i}$ $\in$ $\mathcal{S'}$, and 0 otherwise. The problem can now be formulated as follows.
\begin{problem}
	\label{prob:optgeo}
	The Optimal Multicast Problem
	\begin{alignat}{2}
	& {\text{minimize}} \quad
	& & \sum_{i=1}^{K}{x_{i}d_{i}}  \label{eqn:optgeo-obj} \\
	& \text{subject to} \quad & & \bigcup S_j = \mathcal{U} \quad \forall \quad S_j \in \mathcal{S'}  \label{eqn:optgeo-con1} \\
	& \text{where} & & S_i \in \mathcal{S'} \text{ if } x_i = 1 \nonumber
	\end{alignat}
\end{problem}
In the objective (Equation~\ref{eqn:optgeo-obj}), the delay $d_i$ of each set is the cost (Equation ~\ref{eqn:subsetDelay}) of broadcasting data to all nodes in that set. Equation~\ref{eqn:optgeo-con1} stipulates that each node has to be in at least one set (i.e., every node receives the broadcast data).

\subsection{Solution: Set Cover}\label{setCover}
In this subsection, we translate the Optimal Multicast Problem into a weighted set cover problem. Formally, the minimum weighted set cover problem is as follows. Given a universe $\mathcal{U}$ of $N$ elements, and a collection $\mathcal{S} = \{S_1, S_2, \dots, S_K\}$ of sets whose elements are in $\mathcal{U}$, where each set $S_i$ is assigned a weight $w_i$, the objective is to find a subset $\mathcal{S'}$ of $\mathcal{S}$ with minimum total weight such that each element in $\mathcal{U}$ exists in at least one set in $\mathcal{S'}$ (i.e., all elements are ``covered''). We can easily see that the Optimal Multicast Problem in~\ref{brodProb} is equivalent to the minimum weighted set cover problem. The set $\mathcal{U}$ of $N$ nodes maps to the universe $\mathcal{U}$ of elements, the delay $d_i$ associated with each subset $S_i$ maps to the weight $w_i$ in the weighted set cover problem. The CPLEX Optimization Studio was used to solve this integer programming instance of the Optimal Multicast Problem.

\subsection{Solution: Heuristic}\label{Heuristic}
Due to the computational complexity (demonstrated in Section~\ref{perfEval}) of solving the above integer program, We provide a greedy heuristic. The heuristic builds sets by greedily comparing the cost of broadcasting to the cost of performing multiple unicasts to a pair of adjacent nodes. The delay $d_{i,i+1}$ associated with broadcasting to a pair of adjacent nodes $n_{i}$ and $n_{i+1}$ is defined as the weight $d$ (Equation~\ref{eqn:subsetDelay}) of a set $S=\{n_i, n_{i+1}\}$. Similarly, the delay $d_i$ associated with unicasting to a node $n_i$ is defined as $P/R_b(\theta_i)$.

\newlength{\oldtextfloatsep}\setlength{\oldtextfloatsep}{\textfloatsep}
\setlength{\textfloatsep}{0pt}
\begin{algorithm}[t]
\small
\caption{Greedy Local Optimum Heuristic}
\label{alg:heuristic}
\KwIn{Location ($x_{i}$, $y_{i}$) for nodes $n_{1}$ \textbf{to} $n_{N}$, P, $d_{al}$ }
\KwOut{Sets containing nodes in multicast group}

\For{$i \gets 1$ \textbf{to} $N$\label{orientStart}} { 
	$\text{$\phi_i$} \gets tan (\frac{y_{i}-y_{0}}{x_{i}-x_{0}}) $
}

Sort nodes  in descending order of $\phi_i$ 

$j \gets 1$  

$S_{j} \gets n_{1}$ 

\For{$i \gets 1$ \textbf{to} $N-1$ } {
	\If{$d_{i,i+1} < d_{i}+ d_{i+1}+ d_{al}$} {
    		
    		$S_{j} \gets S_{j}\bigcup n_{i+1}$ 
 	 }
	\Else { 
    	$j \gets j+1$ 
    	
    	$S_{j} \gets n_{i+1}$
	}
}
 
\end{algorithm}
\setlength{\textfloatsep}{\oldtextfloatsep}

The heuristic presented in Algorithm~\ref{alg:heuristic} takes as input the node coordinates, alignment delay $d_{al}$ and packet size $P$. $x_{0}$ and $y_{0}$ are the coordinates of the sender. In lines 1 to 3, the sender sorts the receiving nodes in order of decreasing azimuth $\phi_i$ from the origin (where the sender is located). A set is then created and the first node in the array of sorted nodes is placed in it (lines 4,5). In lines 6 to 11, the algorithm compares the delay associated with broadcasting to a pair of adjacent nodes to that of multiple unicast with alignment delay $d_{al}$ accounted for. If broadcast is cheaper, both nodes are placed in the same set $S_j$ (line 8). On the other hand, if unicast is cheaper, each node in the pair is placed in separate sets (lines 9-11). On the next iteration, the same process is repeated with the next node in the sorted array either being placed in the set containing its neighbor or a new set, depending on whether broadcast or multiple unicast is cheaper. When a broadcast to $N$ nodes is required, $N-1$ such comparisons are made, meaning that the heuristic runs in $O(N)$ time.

\section{Performance Evaluation}\label{perfEval}
In this section we analyze and compare simulation results for various schemes: na\"{i}ve broadcast (FSO BCast) where $\theta$ includes all nodes, multiple unicast(MU) which performs $N$ unicast transmissions, our proposed heuristic (Algorithm~\ref{alg:heuristic}) and the integer programming based set cover solution (Set Cover). We evaluated the performance using delay, throughput and computation time (on a Intel Core i7-4790 based PC) as metrics, with a Java based simulator. Set Cover was implemented using the CPLEX Optimization Studio within the simulator. Each data point is the result of an average over 5000 random node locations. The simulations were performed using an RF range of 150 m. The maximum divergence angle was set to 90 degrees. The parameters we use for the analysis are total data size $P$, GPS error, alignment delay $d_{al}$ and number of nodes $N$. The realistic default values (and ranges) used are: $P$=100 GB (20-180 GB), a GPS error of 3 m (1-5 m), $d_{al}$=2 s (1-3 s) and $N$=15 (10-25). In addition to these parameters, we used a FSO transmit power $P_t$=13 dBm, a wavelength $c/f$=1550  nm, a receiver diameter $D$=12 mm, a photodetector sensitivity $N_b$=0.1875 photons/bit, and 867 Mbps SISO RF links.
\begin{figure}[t]
	\centering
	\subfloat{\includegraphics[width=0.8\columnwidth]{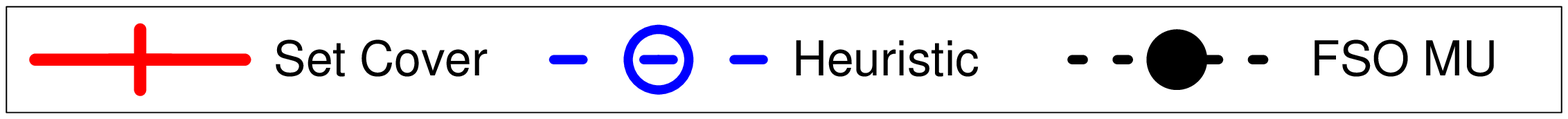}}\vspace*{-0.5em}
	
	\addtocounter{subfigure}{-1}
	\subfloat[]{\includegraphics[width=0.19\paperwidth]{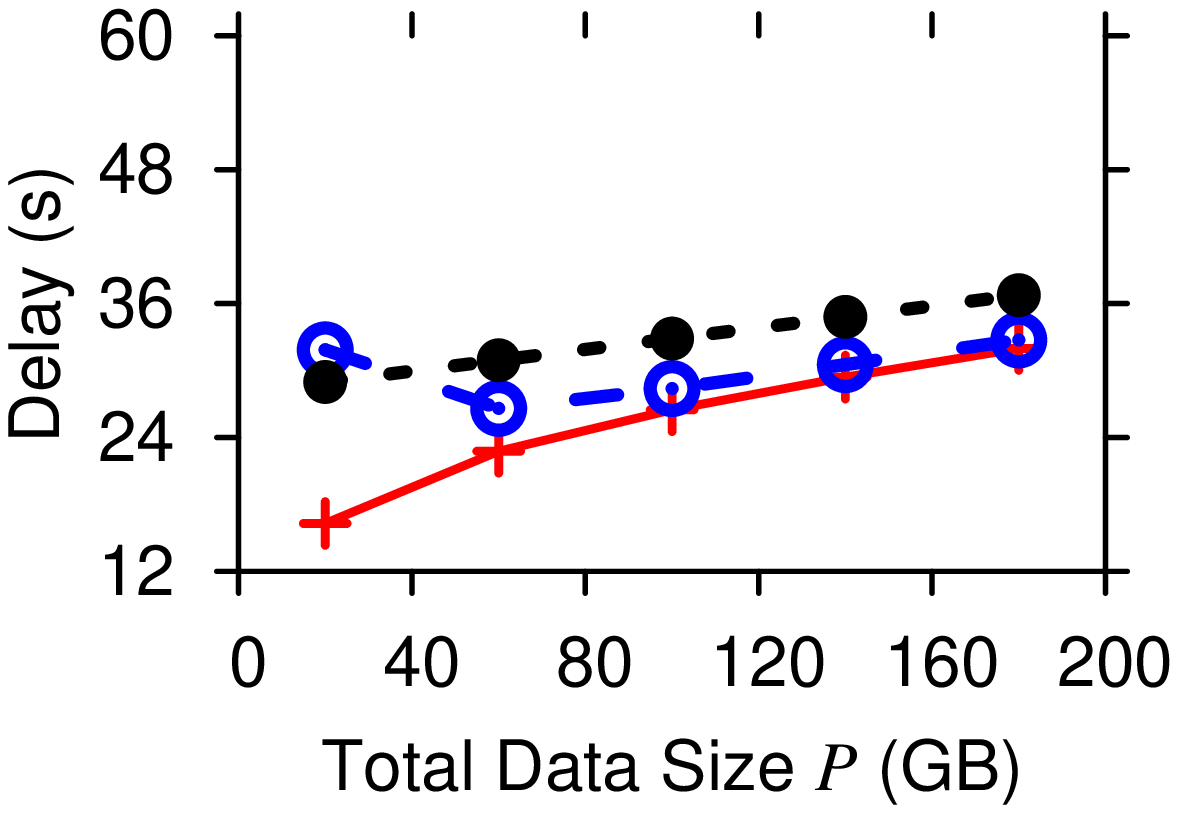}\label{fig:delayData}}\hspace*{1em}
	\subfloat[]{\includegraphics[width=0.19\paperwidth]{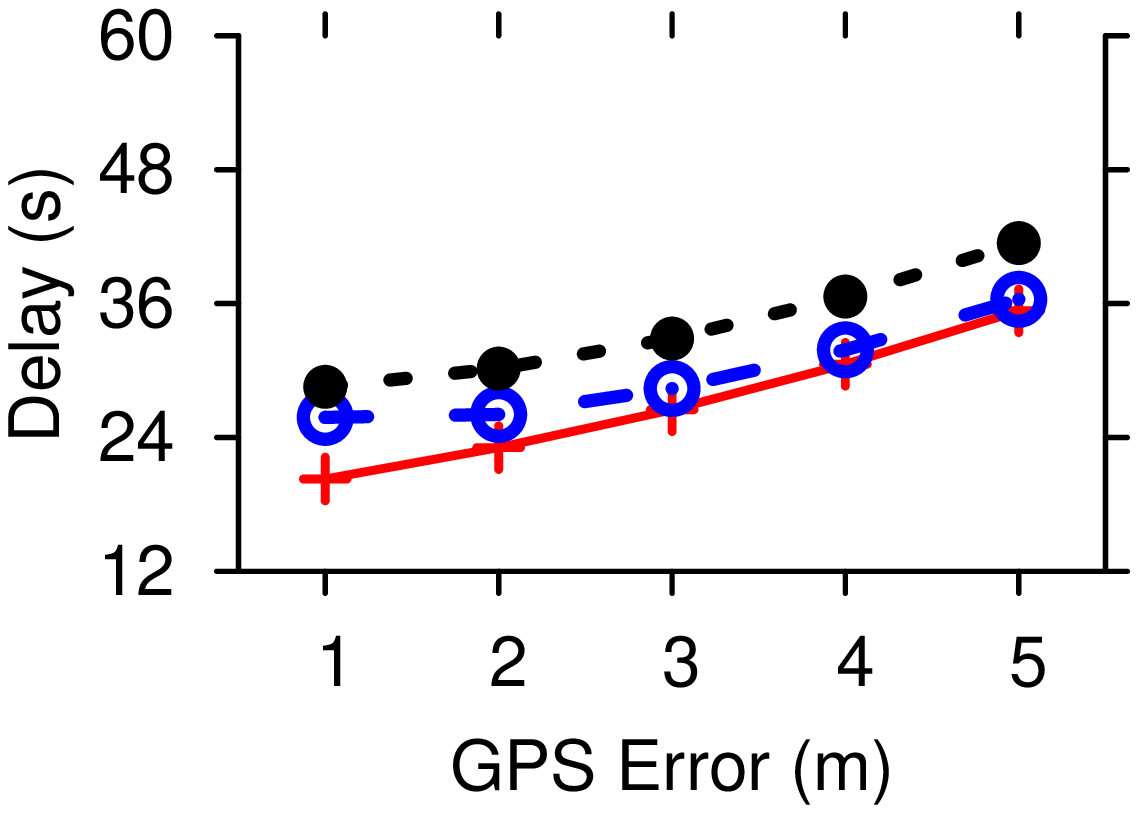}\label{fig:delayGPS}}
	
	\subfloat[]{\includegraphics[width=0.19\paperwidth]{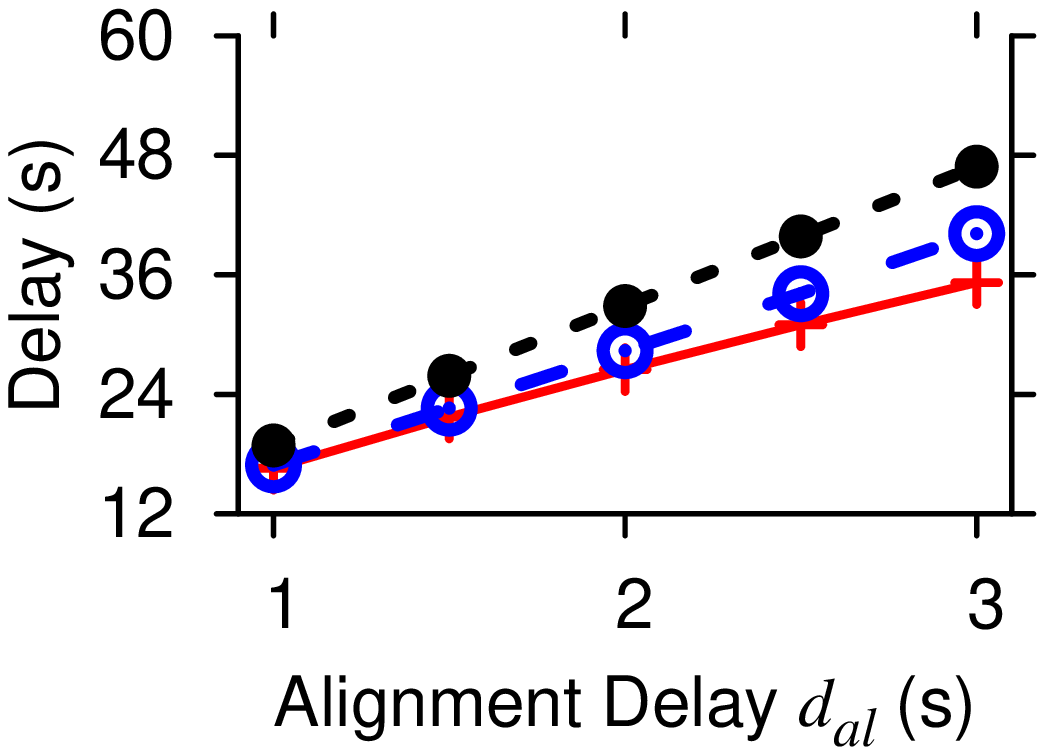}\label{fig:delayAlign}}\hspace*{1em}
	\subfloat[]{\includegraphics[width=0.19\paperwidth]{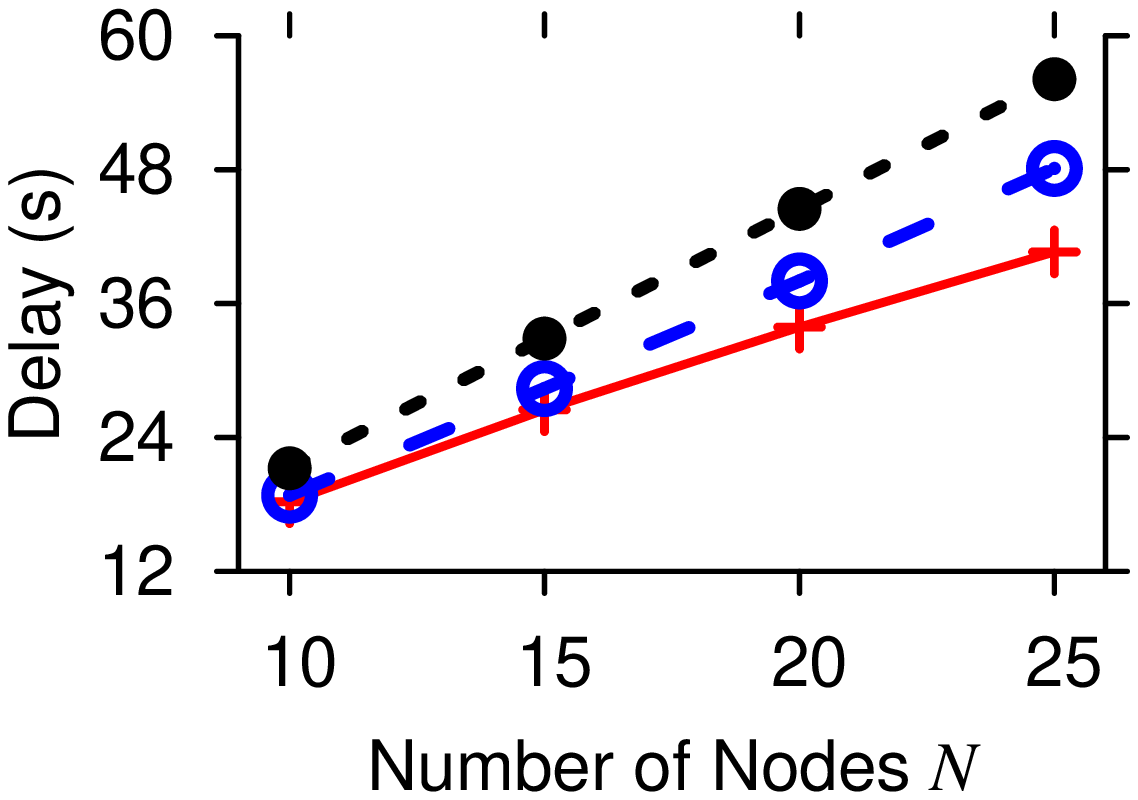}\label{fig:delayNodes}}
	\caption{Delivery delay versus various network parameters. (a) Effect of $P$ on delay for $P$=20-180 GB, (b) Effect of GPS error on delay for GPS error=1-5 m, (c) Effect of $d_{al}$ on delay for $d_{al}$=1-3s, (d) Effect of $N$ on delay for $N$=10-25.}
	\label{fig:delay}
\end{figure}
 
\subsection{Data Delivery Delay}
Delivery delay is defined as the total time it takes to broadcast data to all nodes. In Figure~\ref{fig:delay}, we see that in general, Set Cover offers the best (least) delay followed by our solution (Heuristic), multiple unicast (FSO MU) and broadcast (FSO BCast). Results for FSO BCast are not presented since the delay is very high (due to a large $\theta$) and makes the graphs illegible. The CPLEX-based Set Cover result is obtained using branch and bound. Since the Heuristic only finds the local optimal solution for a pair of nodes when forming sets, it does not discover the optimal $\mathcal{S'}$. However, it still performs better than FSO MU, which suffers from $(N-1)d_{al}$ re-alignment delays. In Figure~\ref{fig:delayData}, as data size $P$ increases, total delay also increases. Transmission delay represents a major component of total delay and it is directly proportional to $P$. Therefore, as $P$ increases, transmission delay per set increases, leading to an overall increase in delivery delay. We see in Figure~\ref{fig:delayGPS} that for all schemes, a decrease in GPS accuracy (increase in GPS error) results in an increase in delay. This is because, as GPS error increases, the $\theta$ required to reach a node while accounting for positioning error increases (i.e., the angle between the two tangents to a circle increases with increasing radius). The greater the $\theta$, the lower the $R_b$, resulting in an increase in transmission delay. The total delay increases as alignment delay $d_{al}$ increases (Figure~\ref{fig:delayAlign}). This is fairly obvious, because the longer it takes to realign the transmitter, the longer it takes for the last bit to be received. As the number of nodes increases, we observe in Figure~\ref{fig:delayNodes} that delay increases. To explain this, we note that $|\mathcal{S'}|$ for a scenario with $N'>N$ is greater than or equal to $|\mathcal{S'}|$ for a scenario with $N$ nodes. Also, $\theta_{i}$ for each $S_i$ $\in$ $\mathcal{S'}$ in the scenario with $N'>N$ is greater than or equal to  $\theta_{i}$ for each $S_i$ $\in$ $\mathcal{S'}$ in the scenario with $N$ nodes. This combination of factors explain the increase in delay as the number of nodes increases.
 
\begin{figure}[t]
	\centering
	\subfloat{\includegraphics[width=\columnwidth]{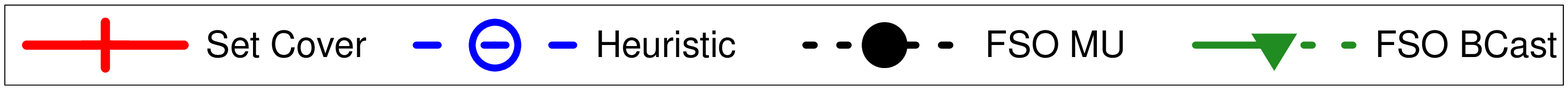}}\vspace*{-0.5em}

	\addtocounter{subfigure}{-1}
	\subfloat[]{\includegraphics[width=0.2\paperwidth]{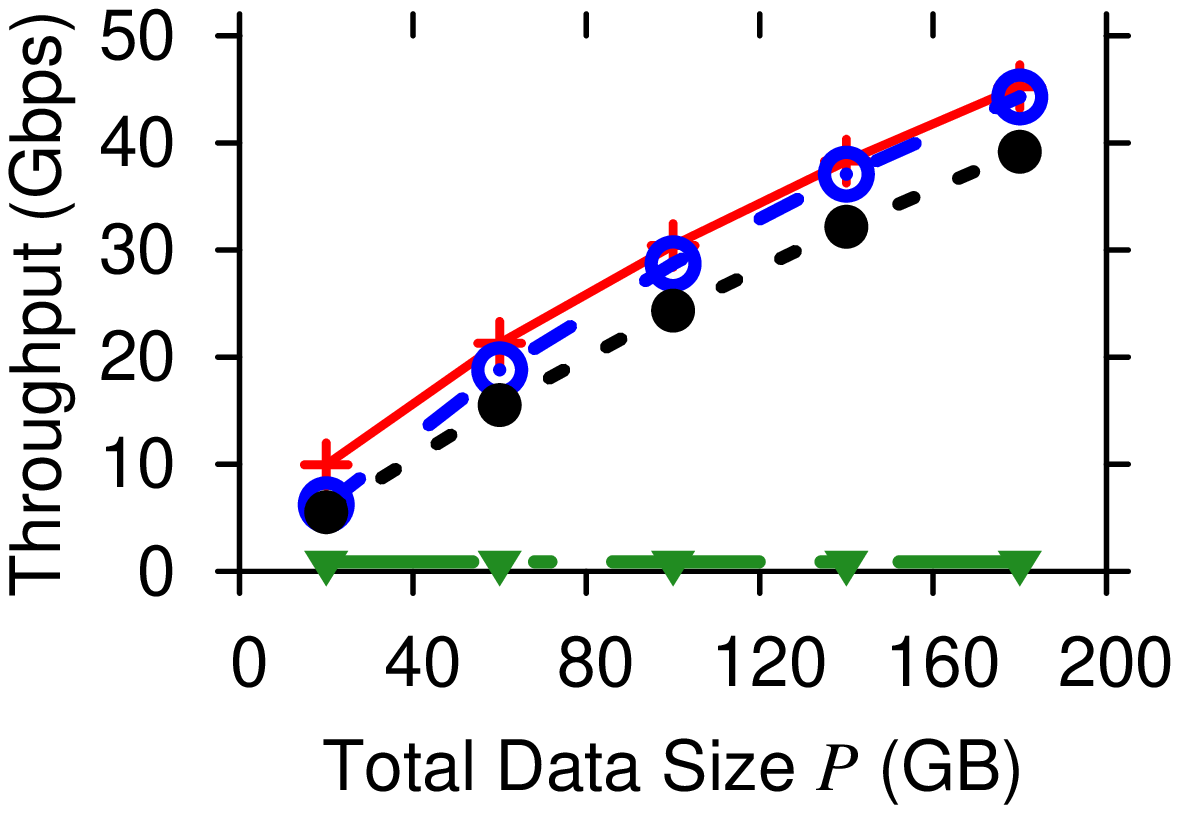}\label{fig:thruputData}}\hspace*{1em}
	\subfloat[]{\includegraphics[width=0.2\paperwidth]{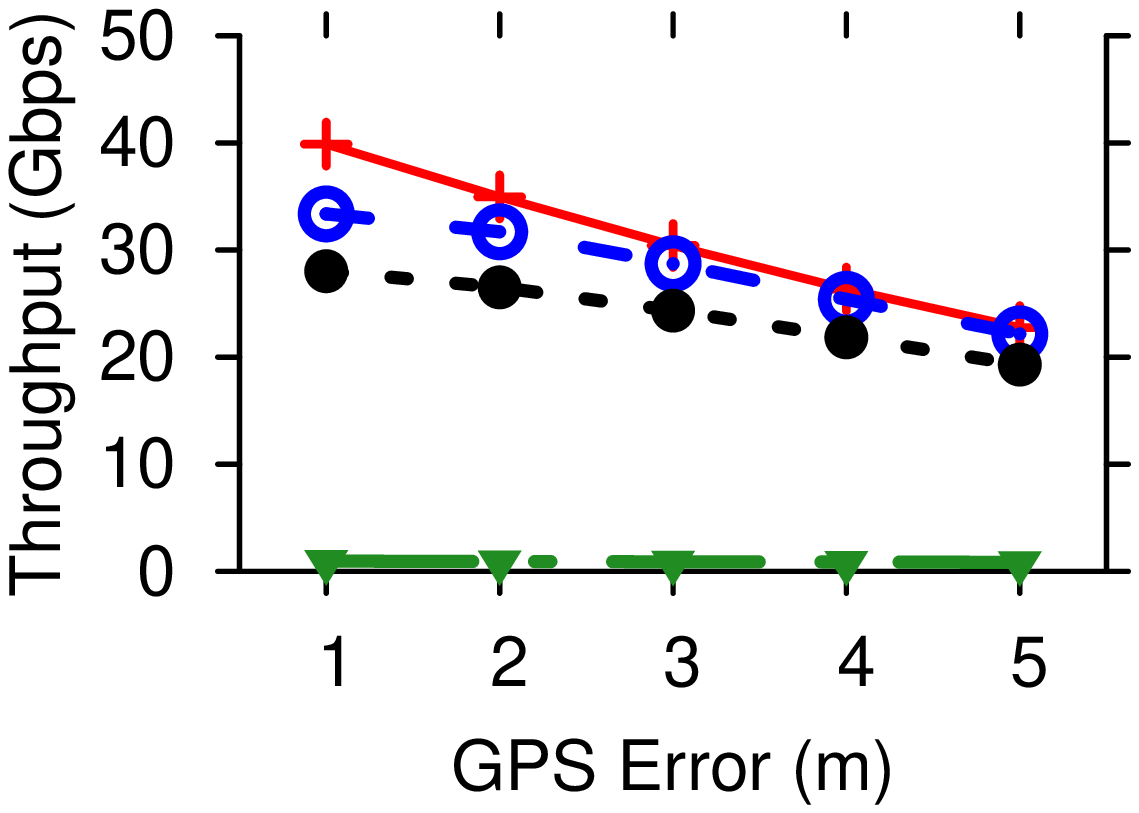}\label{fig:thruputGPS}}
	
	\subfloat[]{\includegraphics[width=0.2\paperwidth]{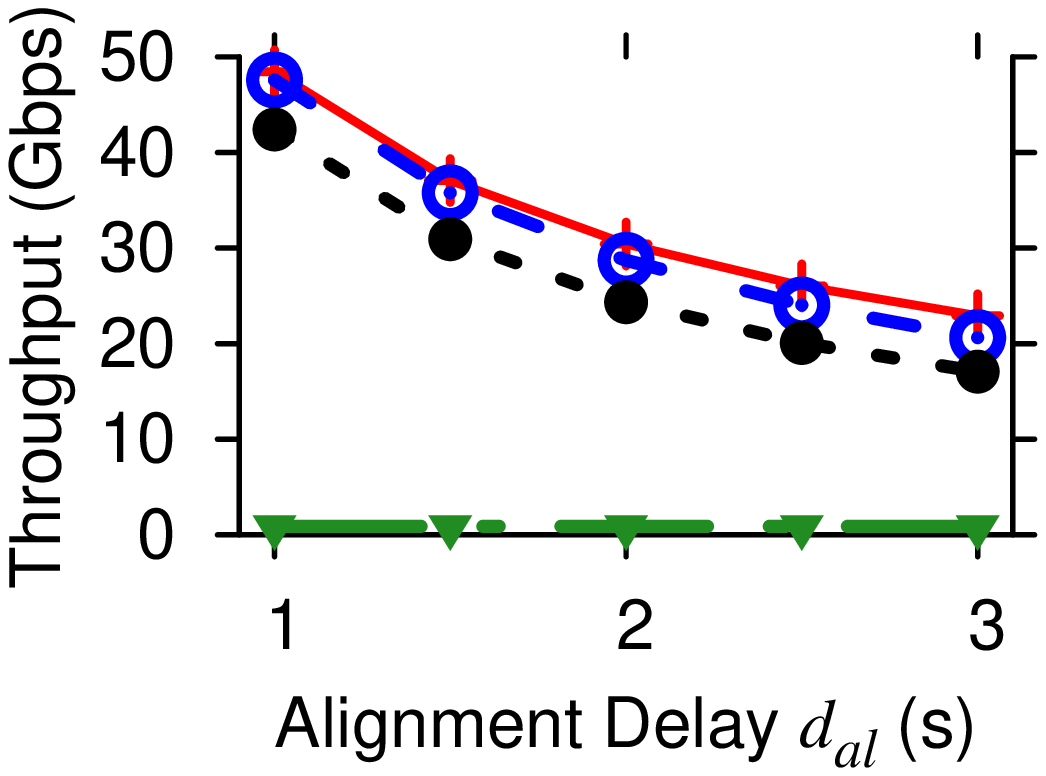}\label{fig:thruputAlign}}\hspace*{1em}
	\subfloat[]{\includegraphics[width=0.2\paperwidth]{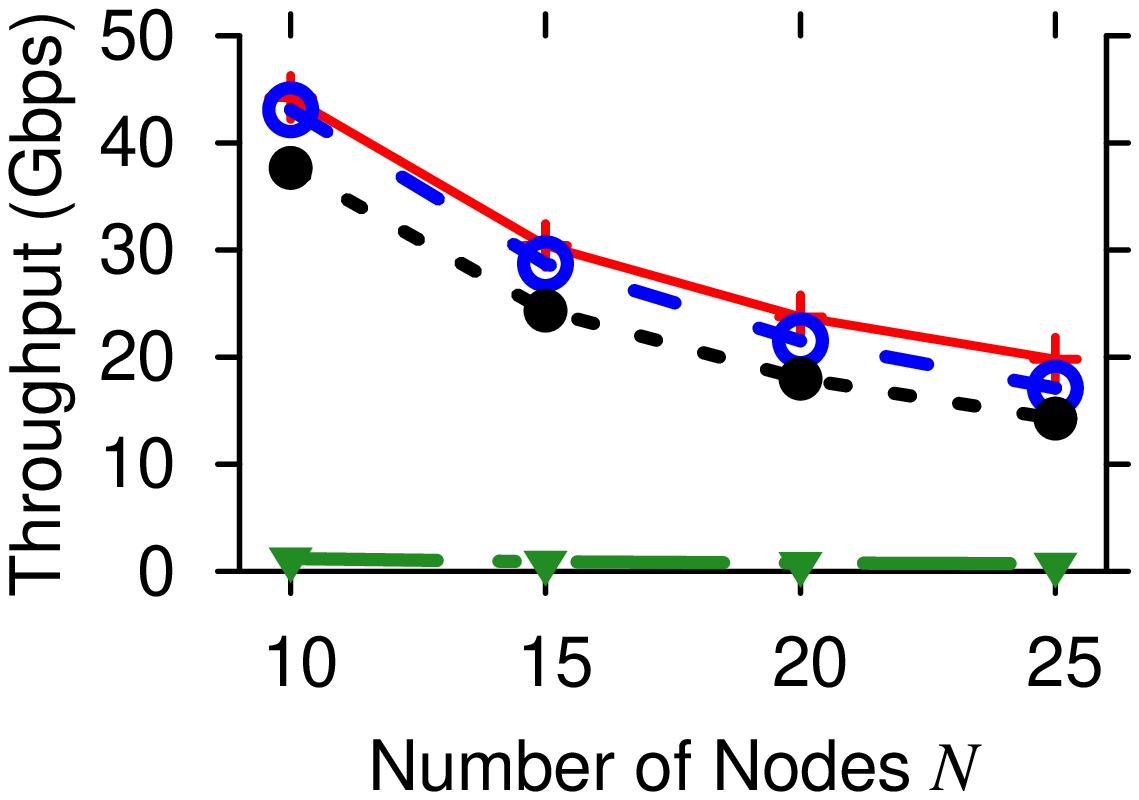}\label{fig:thruputNodes}}
	\caption{Average throughput versus various network parameters. (a) Effect of $P$ on throughput for $P$=20-180 GB, (b) Effect of GPS error on throughput for GPS error=1-5 m, (c) Effect of $d_{al}$ on throughput for $d_{al}$=1-3 s, (d) Effect of $N$ on throughput for $N$=10-25.}
	\label{fig:thruput}
\end{figure}
 
 \begin{figure}[t]
 	\centering
 	\subfloat{\includegraphics[width=\columnwidth]{legend}}\vspace*{-0.5em}
 	
 	\addtocounter{subfigure}{-1}
 	\subfloat[]{\includegraphics[width=0.2\paperwidth]{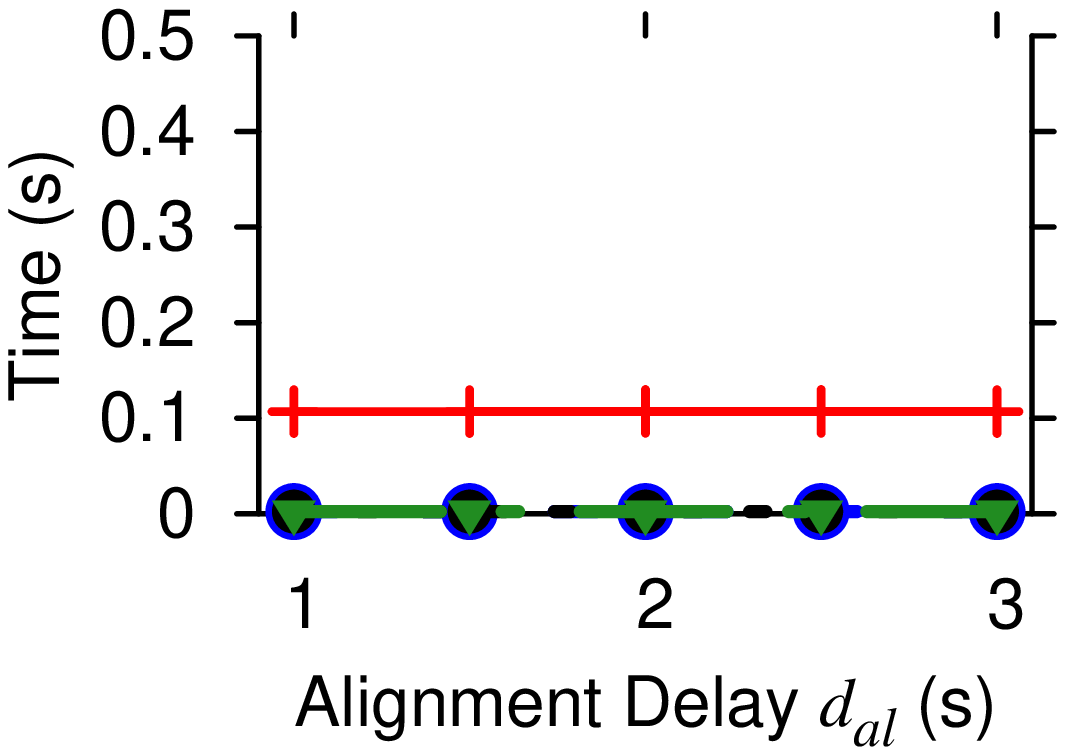}\label{fig:compAlign}}\hspace*{1em}
 	\subfloat[]{\includegraphics[width=0.2\paperwidth]{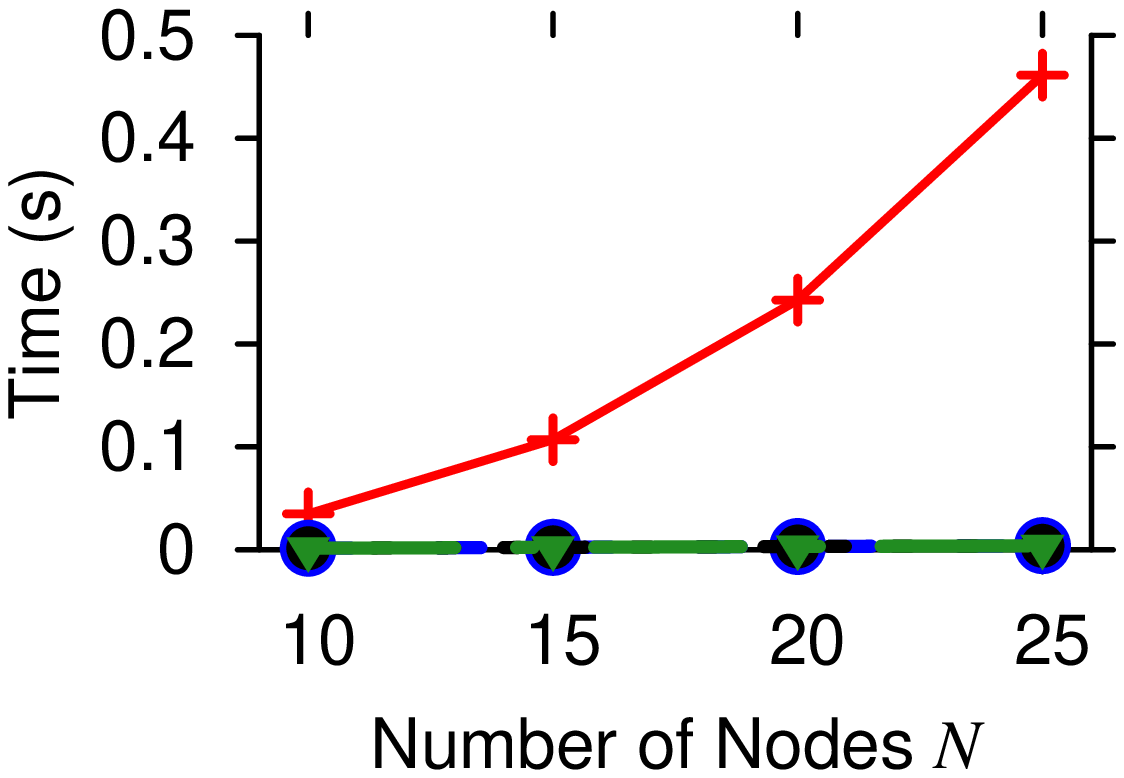}\label{fig:compNodes}}
 	\caption{Computation time versus various network parameters. (a) Effect of $d_{al}$ on computation time for $d_{al}$=1-3 s, (b) Effect of $N$ on computation time for $N$=10-25. The first result (a) is identical to results for variations in $P$ and GPS error (not shown).}
 	\label{fig:compTime}
 \end{figure}

\subsection{Average Throughput}
We define average throughput to be aggregate throughput (total data transferred/total time taken) divided by the number of nodes. The effect of various parameters on average throughput is shown in Figure~\ref{fig:thruput}. Again, Set cover solution offers the best (highest) throughput followed by Heuristic, FSO MU, and FSO BCast. The throughput associated with FSO BCast is significantly lower than the other schemes, since $\theta$ is chosen such that all nodes are reached in exactly one transmission. This leads to significantly high transmission delays, and hence low throughput. As data size $P$ increases, throughput also increases (Figure~\ref{fig:thruputData}). One might expect that for each solution, the throughput stays constant for different values of $P$ (as is the case in broadcast). However, the total delivery delay depends not only on $P$, but also the non-zero alignment delay $d_{al}$ which is constant and independent of $P$. An increase in GPS error leads to a reduction in throughput for all schemes as shown in Figure~\ref{fig:thruputGPS}. This is because in accounting for GPS error, the minimum divergence angle per node becomes larger as previously explained. Hence, the minimum divergence angle $\theta$ required to reach all nodes in a set is at least equal to that when the GPS error is 1 m. The greater $\theta$ is, the greater the transmission delay, hence the reduction in throughput. For all schemes, it is quite obvious that an increase in alignment delay $d_{al}$ results in a decrease in throughput as shown in Figure~\ref{fig:thruputAlign}. This is because, as alignment delay increases, the time it takes for data to reach all nodes increases, resulting in a lower throughput. With an increase in node density, throughput decreases as highlighted in Figure~\ref{fig:thruputNodes}. In explaining this observation, we note that, for a scenario with $N'>N$, $|\mathcal{S'}|$  is greater than or equal to $|\mathcal{S'}|$ for a scenario with $N$ nodes. Also,for each $S_i$ $\in$ $\mathcal{S'}$, $\theta_{i}$ in the scenario with $N'>N$ is greater than or equal to  $\theta_{i}$ for each $S_i$ $\in$ $\mathcal{S'}$ in the scenario with $N$ nodes. We illustrate this using a FSO MU example. Supposing a scenario contains 25 nodes ($|\mathcal{S'}|$=25) and another scenario contains 10 nodes ($|\mathcal{S'}|$=10), we see that the scenario with 25 requires a greater number of realignments since $|\mathcal{S'}|$ is greater. The total delay therefore increases, leading to a reduction in throughput.    

\subsection{Computation time}
We present the graphs of computation time versus the various network parameters in  Figure~\ref{fig:compTime}. The graphs for the effects of data size and GPS error on computation time were identical to Figure~\ref{fig:compAlign}. In Figure~\ref{fig:compAlign}, we observe that set cover takes the most time to find a solution (about 0.1 sec). FSO BCast and FSO MU require the least amount of time to find a solution since less computation is required. For example FSO BCast just needs to find the minimum divergence angle required for data to reach all nodes in a single transmission. The heuristic takes slightly more time to find a solution compared to both FSO MU and FSO BCast. As explained in~\ref{Heuristic}, a solution is found in $O(N)$ time. In obtaining the exact solution, the integer programming implementation takes the most time. This is because CPLEX uses branch and bound to obtain the optimal solution. With branch and bound, a sequence of subproblems are formed which converge to the optimal solution. The result in Figure~\ref{fig:compNodes} is particularly interesting. As the number of nodes increases, the number of sets $K$ in the integer program increases quadratically. Since the integer program formulation NP-hard, the run time increases exponentially with $N$. 

\section{Conclusion}
Arbitrarily increasing the beam divergence angle of an optical link leads to low data rates, hence broadcasting becomes inefficient. We have presented an optimal multicasting algorithm for broadcast over FSO links in hybrid RF/FSO ad hoc networks. We showed that this problem is an abstraction of the minimum weighted Set Cover problem which is known to be NP-hard. We developed a less computationally intensive greedy heuristic that takes approximately 5\% of the time as the Set Cover solution, but achieves 95\% of the throughput. Our solution performs better than both direct broadcast and multiple unicast solutions.

\section{Future Work}
In this work, we limited the divergence angle to 1 quadrant (90 degrees) for practical purposes. We intend to extend our work to include using a maximum divergence angle of 360 degrees for completeness, transmitting through both RF and FSO simultaneously, and evaluating network performance under noisy channel conditions. In addition, we will integrate our solution into state of art DTN routing protocols.

\tiny
\bibliographystyle{unsrt-abbrv}
\bibliography{qos}

\end{document}